\documentclass[conference,onecolumn]{IEEEtran}

\usepackage{cite}
\usepackage{amsmath,amssymb,amsfonts}
\usepackage{algorithmic}
\usepackage{graphicx}
\usepackage{textcomp}
\usepackage{xcolor}
\usepackage[hidelinks]{hyperref}
\usepackage{amsmath}

\def\BibTeX{{\rm B\kern-.05em{\sc i\kern-.025em b}\kern-.08em
    T\kern-.1667em\lower.7ex\hbox{E}\kern-.125emX}}

\usepackage{amsmath}
\usepackage{graphicx}

\makeatletter

\def\ps@IEEEtitlepagestyle{%
  \def\@oddfoot{\mycopyrightnotice}%
  \def\@evenfoot{}%
}
\def\mycopyrightnotice{%
  {\footnotesize 978-1-6654-9106–8/22/\$31.00~\copyright~2022 IEEE\hfill}% <--- Change here
  \gdef\mycopyrightnotice{}
}

\@ifundefined{showcaptionsetup}{}{
 \PassOptionsToPackage{caption=false}{subfig}}
\usepackage{subfig}
\makeatother

\usepackage{eso-pic}

\title{An Adaptive Strain Estimation Algorithm Using Short Term Cross Correlation Kernels and 1.5D Lateral Search\
}

\begin{document}

%\title{PCONet: A Convolutional Neural Network Architecture to Detect %Polycystic Ovary Syndrome (PCOS) from Ovarian Ultrasound Images\\
%}

\author{\IEEEauthorblockN{Shaiban Ahmed}
\IEEEauthorblockA{\textit{Biomedical Engineering
} \\
\textit{University of Illinois Chicago}\\
Chicago, IL, USA \\
sahme83@uic.edu}
\and

\IEEEauthorblockN{Rasheed Abid}
\IEEEauthorblockA{\textit{Biomedical engineering} \\
\textit{Illinois Institute of Technology}\\
Chicago, IL, USA \\
r.abid94.bogra@gmail.com}

\and
\IEEEauthorblockN{S. Kaisar Alam}
\IEEEauthorblockA{\textit{Rutgers University}\\
Newark, NJ,USA \\
kaisar.alam@ieee.org}
}

\maketitle

\begin{abstract}
Adaptive stretching, where the post compression signal is iteratively stretched to maximize the correlation between the pre and post compression rf echo frames, has demonstrated superior performance compared to gradient based methods. At higher levels of applied strain however, adaptive stretching suffers from decorrelation noise and the image quality deteriorates significantly. Reducing the size of correlation windows have previously showed to enhance the performance in a speckle tracking algorithm but a correlation filter was required to prevent peak hopping errors. In this paper, we present a novel strain estimation algorithm which utilizes an array of overlapping short term cross correlation kernels which are about one-fourth the size of a typical large kernel, to implement an adaptive stretching algorithm. Our method does not require any supplementary correlation filter to prevent false peak errors. Additionally, a lateral search is incorporated using 1.5D algorithm to account for the mechanically induced lateral shift. To validate the efficacy of our proposed method we analyzed the results using simulation and in-vivo data of breast tumors. Our proposed method demonstrated a superior performance compared to classical adaptive stretching algorithm in both qualitative and quantitative assessment. Strain SNRe, CNRe and image resolution are found to improve significantly. Lesion's shape and boundary are more clearly depicted. The results of improvement are clearly evident at higher levels of applied strain.

\end{abstract}


\section{Introduction}
Medical imaging is a powerful means of diagnosis of various kind of diseases in the human body  ~\cite{Garra1997-nv}, ~\cite{Konofagou2002-nr},  ~\cite{Emelianov1995-vu},~\cite{Hosain2022-kr}, ~\cite{Hosain2022-mg}. Specifically, strain Elastography, which has garnered wide scale popularity ~\cite{Ophir1991-rh},~\cite{Cespedes1993-rv},~\cite{Ophir1996-wi},~\cite{Ophir2002-vy}, ~\cite{Zheng2021-el} over the years due to its non-invasive and inexpensive nature, is primarily an imaging procedure that can map the elastic features of biological tissues and provide an extensive visual and quantitative analysis of the discriminant tissues properties and can be used to assess intricate tissue features ~\cite{Sigrist2017-sv}. As an imaging technique, Elastography has been used in numerous clinical applications such as diagnosis of breast ~\cite{Garra1997-nv},~\cite{7835370},~\cite{Kabir2022-ze}, myocardium imaging ~\cite{Konofagou2002-nr}, renal pathology ~\cite{Emelianov1995-vu} etc. Pathology and in some cases physiological phenomena change the stiffness of many tissues and this change can be detected by manual palpation, a method that has been used for millennia as a diagnostic tool.  However, free hand palpation based diagnosis is limited to detection of abnormal tissues having a significant difference in stiffness compared to their surroundings. Also palpation is subjective and clinician dependent making independent confirmation of findings difficult. Elastography can recreate palpation like ability by using advanced computational algorithm to detect and accurately analyze the response from an applied stimuli and thus, examine tissue features with more efficiency and accuracy. Quasi-static strain imaging techniques based on Elastography are compression based methods where an external excitation or stimulus is generated by mechanically compressing the tissue surface by using ultrasound transducers ~\cite{285465}. In Strain Elastography, ultrasound echoes are recorded before and after applying the mechanical compression. These pre and post deformation signals are then used to estimate strain. Numerous correlation based algorithms have been presented to estimate strain from the pre and post compression signals. These fall into two groups: a) Gradient based approaches ~\cite{Ophir1991-rh},~\cite{Cespedes1993-rv},~\cite{Ophir1996-wi},~\cite{Ophir2002-vy}, ~\cite{1703750}, ~\cite{1703750} and b) Direct strain estimators ~\cite{Ophir1999-fl},~\cite{Varghese2000-qn},~\cite{Alam2004-gf}. In gradient based methods, strain is computed by calculating the displacement derivative. Where, displacement due to the applied mechanical compression can be calculated from time-delay or phase shift. Time delay can be estimated by computing cross-correlation ~\cite{Ophir1991-rh},~\cite{Cespedes1993-rv},~\cite{Ophir1996-wi},~\cite{Ophir2002-vy} of pre and post compression signals. Alternatively, phase shift can be estimated from phase domain multiplication ~\cite{234286},~\cite{796111},~\cite{584292}. However such methods are prone to noise amplification. Stretching the post compression signal using a global stretch factor ~\cite{Cespedes1993-gl}, ~\cite{Alam1997-ph} prior to correlation computation has previously resulted in noise reduction but only appeared to be effective at low strain ~\cite{Alam1997-ph}. Median filtering or least square based techniques such as linear regression ~\cite{Kallel1997-hq} or smoothing spline ~\cite{Alam2010-gf} can also be applied to enhance performance. But the applied strain itself induces decorrelation noise which can compound with higher levels of applied strain. Direct strain imaging techniques have presented more robust performance compared to gradient based methods where the strain is computed directly from pre and post deformation echo signals. Adaptive strain estimation ~\cite{Alam1998} or an adaptive spectral strain estimation ~\cite{Alam2004-gf} is a direct strain estimation approaching where a local stretch factor is used to stretch the post compression signal to improve cross correlation accuracy for strain estimation. Adaptive stretching of the post compression signal instead of global stretching leads to more accurate strain estimates because stretching the post compression signal using one global stretch factor can lead to inaccurate estimates as the tissue displacement is not homogeneous in all regions. However, the classical adaptive strain estimation algorithm was implemented using cross correlation windows having length much longer than the autocorrelation width of the signal. Strain maps generated by such large windows are found to have low spatial resolution and are prone to decorrelation noise and low SNR at higher levels of applied strain. Shorter cross correlation windows were used previously ~\cite{741427} and appeared to enhance the performance of a speckle tracking algorithm where the strain was estimated from a difference estimate but a correlation filter was required to reduce false peak errors and to achieve high SNR values. 
In this paper we investigate the effects of short term cross correlation in a direct strain estimation approach by implementing adaptive stretching algorithm using short term correlation kernels. Strain is estimated from a mean short term cross correlation function rather than using an additional correlation filter to eliminate peak hopping errors. A lateral search is also conducted to search for the lateral shift and have better strain estimates for high strain values. The performance of the proposed method has been thoroughly evaluated by using 2-D Finite Element Model simulated data as well as in-vivo data of breast tumors.

%%%%%%%%%%%%%%%%%%%%%%%%%%%%%%%%%%%%%%%%%%%%%%%%%%%%%%%%%%%%%%%%%%%%%%%%%%%%%%%%%%%%%%%%%%%%%%%%%%%%%%%%%%%%%%%%%%%%%%%%%%%%%%%%%%%%%%%%%%%%%%%%%%%%%%%%
\section{METHOD}
\subsection{Theory of Ultrasound Signal Model and Adaptive Estimator of Strain: }
For the initial analysis a simple one-dimensional (1D) model is developed below to exhibit ultrasound back-scattered radio-frequency (RF) signals. 
We can represent the pre and post deformation echo signals by ~\cite{Alam1998}:

\begin{equation}
\begin{aligned}
r_1(t) &=s_1(t)+n_1(t) \\
\text { or, } r_1(t) &=s(t) * p(t)+n_1(t) \\
\text { or, } r_2(t) &=s_2(t)+n_2(t) \\
\text { or, } r_2(t)=s\left(\frac{t}{a}-t_o\right) * p(t)+n_2(t)
\end{aligned}
\end{equation}

Here, $r_1(t)$ and $r_2(t)$ are the pre and post compression signals respectively; s(t) stands for the effective rf backscatter distribution function in 1-D model; p(t)  is the impulse response of the ultrasonic system;  $n_1(t)$ and $n_2(t)$ are induced uncorrelated random noise processes and the symbol ‘$*$’ represents convolution. For conventional elastography, the applied strain "$\varepsilon$" is significantly smaller than 1. Hence the constant parameter, "a" will be close to unity. Mathematically:
$$
a=1-\varepsilon \cong 1
$$
(2) $\quad[$ when $\varepsilon \ll 1]$

The induced displacement between the pre and post compression signals is depth dependent making them jointly non-stationary. Also, for 1D model, the displacement is assumed to be only in the axial direction. The cross correlation function between $r_1$ (t) and $r_2 (t)$ computed at time $t=t_o$ can be expressed using the equation ~\cite{741427}:

\begin{equation}
\left(r_1(t) \star r_2(t)\right)(\tau)=\hat{R}\left(t_o, t_o+\tau\right)=\frac{1}{T} \int_{t_o-\frac{T}{2}}^{t_o+\frac{T}{2}} r_1(t) r_2(t+\tau) d t
\end{equation}

Here, $"*"$ denotes cross correlation between the two signals; $\tau$ is the induced displacement or correlation lag; T stands for the correlation window or kernel length. The given time domain functions can be converted to their corresponding spatial functions (function of their spatial position) by simply using the transformation equation x = $tc/2$, where x is the distance from the transducer, c is the propagation speed of sound in the tissue medium ~\cite{Alam1998}. Correlation based techniques are widely used for estimating strain where the time delay or displacement is computed by locating the position of peak cross correlation co-efficient ~\cite{285463},~\cite{Bilgen1996-kt},~\cite{Cao2017-dj}. Such correlation based techniques suffer from decorrelation noise which compounds with the increment of applied strain ~\cite{Alam1997-ph},~\cite{Bilgen1996-kt},~\cite{Cao2017-dj}. In order to reduce the effect of applied strain, stretching the post deformation signal temporally prior to correlation computation has previously showed to improve the performance of correlation estimates~\cite{Alam1997-ph},~\cite{1164316}. However, the biological tissues are Heterogeneous in nature and hence, the strain would vary at different window locality. Thus, a varying stretch factor would be ideal instead of a global stretch factor that time stretches the entire post compression A-line at one go. An adaptive algorithm was used previously ~\cite{Alam1998} that uses the stretch factor itself as an estimator of strain. This method uses an adaptively varying local stretch factor $(1/\alpha)$ to stretch the post compression signal temporally. The stretch factor is modified adaptively till the highest cross correlation co-efficient value is achieved. The corresponding value of the stretch factor for which the peak cross correlation co-efficient is obtained $(1/\alpha_{max})$ is then used to estimate strain using the simple relation: 

\begin{equation}
\varepsilon=1-\frac{1}{\alpha_{\max }}
\end{equation}

Let $r_3\left(t\right)$ be the stretched version of $r_2\left(t\right)$ with $(1/\alpha)$ being the stretch factor. Then we can denote this temporally stretched post compression signal as:

\begin{equation}
r_3(t)=r_2(\alpha t)=s_3(t)+n_3(t)=\alpha s\left(\frac{\alpha}{a} t-t_o\right) * p(\alpha t)+n_3(t)
\end{equation}

Now, if we compute the cross correlation between the pre and stretched post compression signal, then the co-efficient of the cross correlation function can be represented by ~\cite{Alam1998}:

\begin{equation}
\widehat{p_{13}}\left(t_o, t_o+\tau\right)=\frac{\frac{1}{T} \int_{t_o-\frac{T}{2}}^{t_o+\frac{T}{2}} r_1(t) r_3(t+\tau) d t}{\sqrt{\frac{1}{T} \int_{t_o-\frac{T}{2}}^{t_o+\frac{T}{2}} r_1^2(t) d t \frac{1}{T} \int_{t_o-\frac{T}{2}}^{t_o+\frac{T}{2}} r_3^2(t+\tau) d t}}
\end{equation}

\begin{equation}
\widehat{p_{13}}\left(t_o, t_o+\tau\right)=\frac{\frac{1}{T} \int_{t_o-\frac{T}{2}}^{t_o+\frac{T}{2}}[s(t) * p(t)]
\left[\alpha s\left\{\frac{\alpha}{a}(t+\tau) -t_o\right\} * p(\alpha t)\right] d t}{\sqrt{\frac{1}{T} \int_{t_0-\frac{T}{2}}^{t_o+\frac{T}{2}}[s(t) * p(t)]^2 d t \frac{1}{T} \int_{t_o-\frac{T}{2}}^{t_o+\frac{T}{2}}\left[\alpha s\left\{\frac{\alpha}{a}(t+\tau)-t_o\right\} * p(\alpha t)\right]^2 d t}}
\end{equation}

\begin{equation}
\widehat{p_{13} \max }=  \frac{\frac{1}{T} \int_{t_o-\frac{T}{2}}^{t_o+\frac{T}{2}}[s(t) * p(t)]\left[s\left\{\frac{\alpha}{a} t\right\} * p(t)\right] d t}{\sqrt{\frac{1}{T} \int_{t_o-\frac{T}{2}}^{t_o+\frac{T}{2}}[s(t) * p(t)]^2 d t \frac{1}{T} \int_{t_o-\frac{T}{2}}^{t_o+\frac{T}{2}}\left[s\left\{\frac{\alpha}{a} t\right\} * p(t)\right]^2 d t}} \leq 1
\end{equation}

Maximum value of this cross correlation co-efficient can be achieved when $\alpha=a$. Selecting an appropriate stretch factor using an adaptive mean, this peak cross-correlation coefficient value is located. Then strain is calculated directly from this stretch factor using Equation 4. However, this classical adaptive stretching method utilizes large correlation kernels which are subject to decorrelation noise at higher strain levels. The resultant strain elastograms hence are predicted to have low SNRe and CNRe values. Also, the large correlation windows can reduce the spatial resolution which in turn is not ideal for imaging the strain map. This problem can be solved by reducing the length of the correlation windows. 

\subsection{Short Term Correlation Kernels:}
Error generated due to variance in displacement or time delay estimation can be limited by reducing the length of correlation kernel. When the length of the correlation kernel becomes smaller, the effect of decorrelation due to applied strain diminishes significantly. However, as the size of these correlation windows become smaller, the normalization terms in equation (6) begin to fluctuate which increases error variance and may results in peak hopping or appearance of false peak along with the true peak in the computed cross correlation function. Previous work using shorter kernels, which have showed to improve the performance in speckle tracking ~\cite{741427} have used a correlation filter to reduce the effect of false peaks. The correlation filter multiplies a number of adjacent correlation function computed at different spatial positions by different weights or filter co-efficients to prevent any false peaks. The accurate length of the filter must be configured to obtain higher SNRe values. Designing such correlation filter can although be computationally intensive. Also, if a number of spatially adjacent correlation functions are filtered, it can lead to reduced spatial resolution. However, usage of such correlation filter can be avoided altogether by using a mean short term cross correlation function as described below.
When a number of adjacent short term cross correlation functions are generated by computing the cross correlation between overlapping short term pre and post compression windows, false peaks are expected to pop up along with the true peak. But the position of this true peak always remain at the same location where the false peaks can rise up at different lag positions. Hence, when all the cross correlation functions are summed together and their mean is computed, the true peak is expected to tower over all the false peaks.

In order to examine this concept, we ran two simulations on a simple 1-D model. The simulations were conducted under an applied strain of 2\%. The outcomes are depicted in Figure 2. The first simulation was conducted using conventional large windows where the cross correlation function was obtained by using a 2.96 mm kernel. The length of the kernel was determined based on empirical analysis and the length (2.96 mm) that produces the highest cross correlation co-efficient was selected. For the second simulation, an array of overlapping short term windows were used. Length of each short term windows was selected to be 0.74 mm based on an empirical analysis where image resolution, strain SNRe and visual inspection was considered. An array of 13 short term kernels with 75\% axial overlapping were computed to cover a length of 2.96 mm to cover the same spatial distance as the large window. By computing cross correlation between the respective short term kernels and search windows, a total of 13 short term cross correlation functions were obtained. Then they were added together and their mean was computed to obtain the mean cross correlation function. The process of generating axially overlapping short term (ST) windows is illustrated in Figure 1. The process to obtain the mean ST cross correlation function is shown in Figure 2(a). Though some false peaks can be observed in the ST cross correlation functions, the mean ST cross correlation function displays only one true peak which in term can be used for locating the displacement or the ideal stretch factor. The mean ST function is plotted along with the cross correlation function obtained by using large windows and the region near the peak is shown in figure 2(b). The result clearly shows that the mean ST cross correlation function computed over the same spatial length sustains a higher and broader correlation peak. As the cross correlation peaks are expected to shift due to tissue heterogeneity, the mean ST cross correlation function produced by adding and averaging the ST cross correlation function produces not only a higher but also a broader correlation peak compared to the cross correlation function obtained by the classical approach. 

\begin{figure}[htp]
    \centering
    \includegraphics[scale=.4]{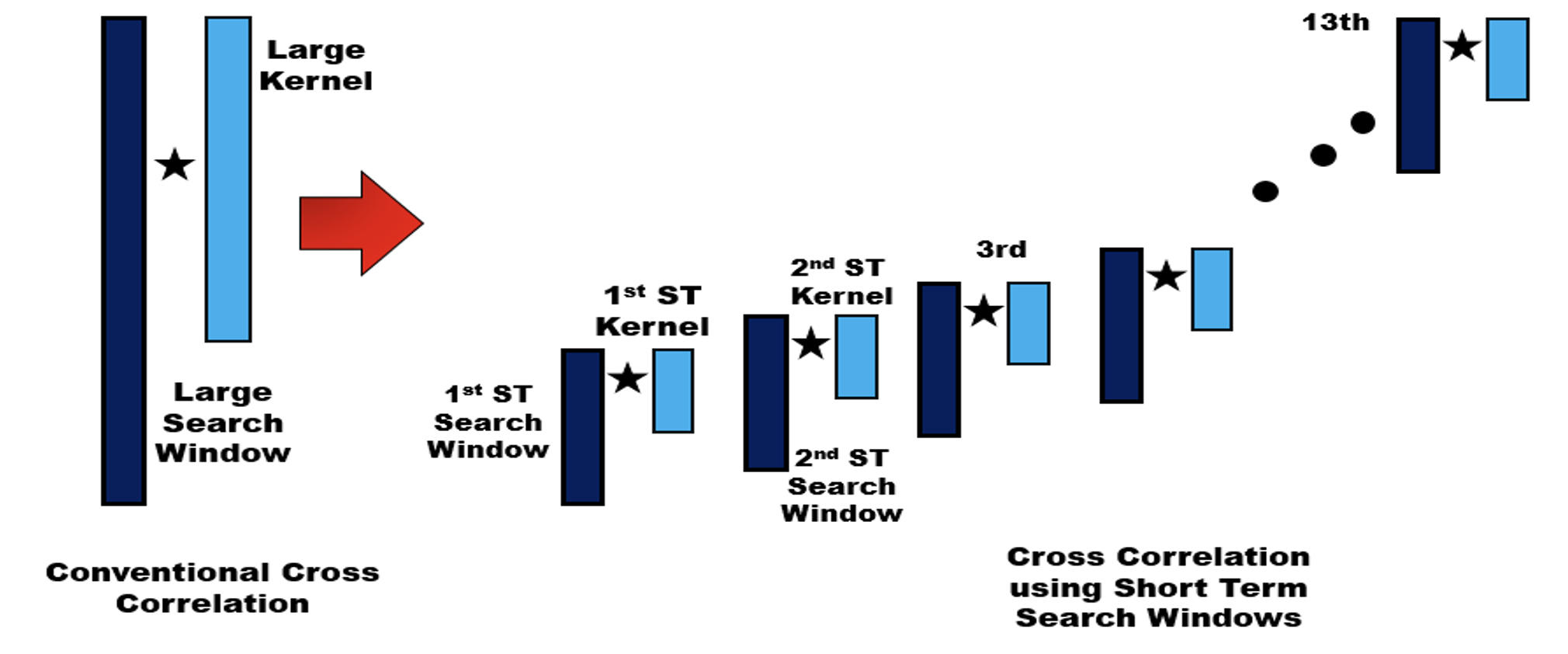}
    \caption{Process of cross correlation using successively generated axially overlapping ST (Short Term) windows is illustrated in contradistinction to conventional correlation using large correlation windows. Here $"\star"$ denotes cross correlation}
    \label{fig:fig1}
\end{figure}

\begin{figure}[htp]
    \centering
    \includegraphics[scale=.4]{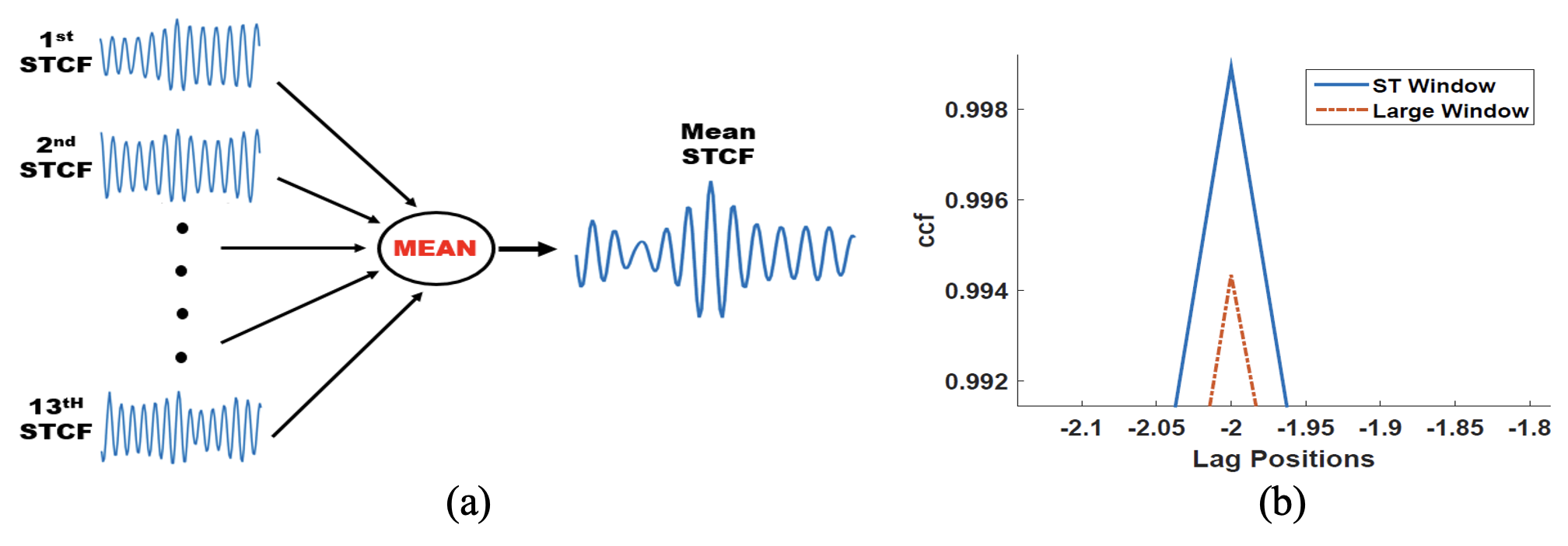}
    \caption{Cross Correlation Functions (a) Mean Short Term Correlation Function generated from succesive Short Term Correlation Functions (STCF) (b) Cross Correlation Co-efficient (ccf) vs Lag Positions to show that the Mean Short Term Corerlation Function generated by using short term windows bears a higher correlation peak compared to the correlation function generated by using conventional Large window}
    \label{fig:fig2}
\end{figure}

\subsection{Lateral Search}

\begin{figure}[htp]
    \centering
    \includegraphics[width=8cm]{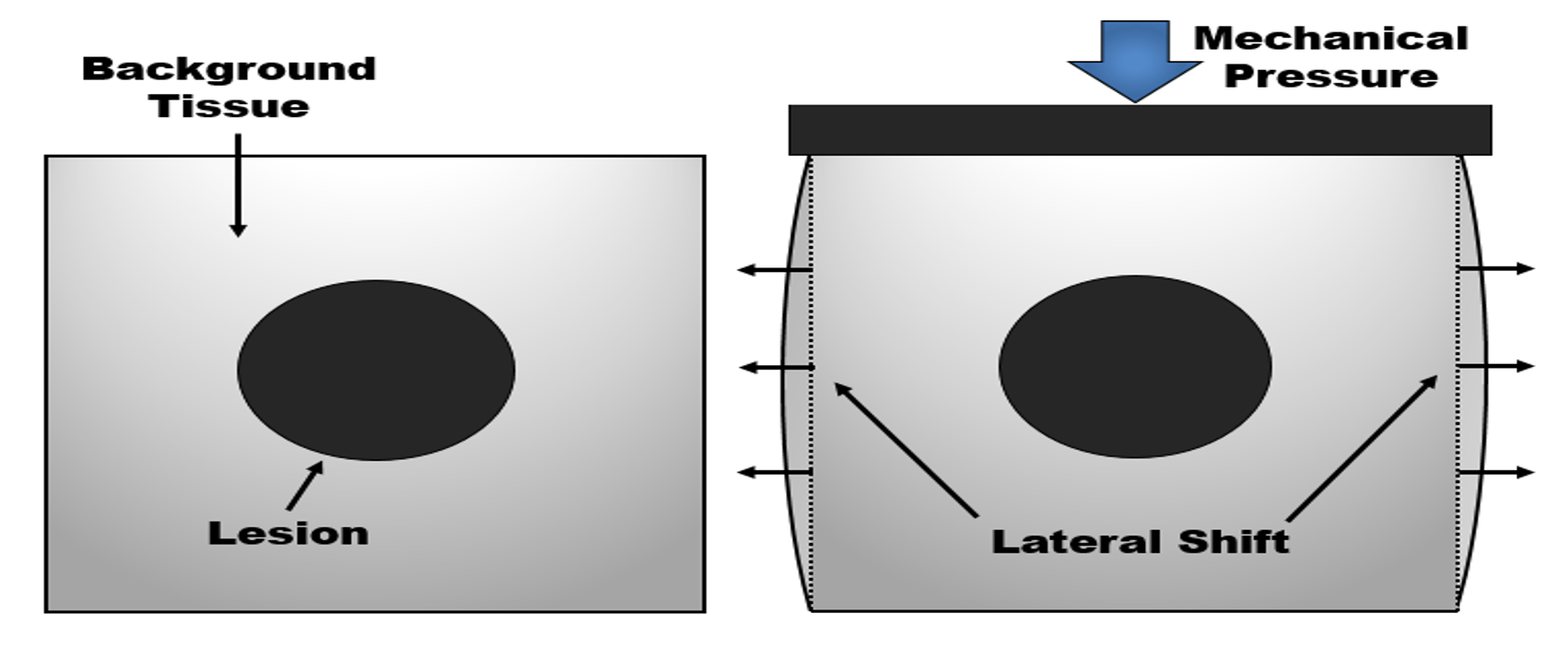}
    \caption{Induced Lateral Shift due to appplied Mechanical compression }
    \label{fig:fig3}
\end{figure}

Our initial method was developed based on a 1-D model where the displacement was assumed to be only along the axial direction which is along the direction of the transmitted ultrasound beam. However, in practical cases, the scenario is quite different. Displacement occurs not only axially but also laterally, which is perpendicular to the direction of transmitted pulse as illustrated in Figure 3. To account for such displacement, a lateral search must be conducted. 2-D search algorithms are developed for this type of analysis where 2-D pre-compression kernels, 2-D post compression search windows and 2-D cross correlation functions are used.  In this paper, in order to reduce computational complexity and to make the processing speed faster, we implement a 1.5D algorithm that utilizes 1-D short term kernels and 2-D short term search windows, with the same dataset presented in ~\cite{7835383}. A preliminary study based on this 1.5D approach was done formerly using simulation phantoms and breasts ~\cite{7835383} where the 1.5D method has comparatively outperformed the classical adaptive stretching algorithm. However, in this preliminary study the 1.5D algorithm produced low spatial resolution strain elastograms at higher strain levels. Additionally the lesion boundaries appeared to be indistinct and speckle noises were apparent around the tumor boundary. In our exploration, we re-implement the 1.5D algorithm using large kernels which are 2.96 mm long and then assess the impact of using short term correlation windows in the 1.5D algorithm.
\begin{figure}[htp]
    \centering
    \includegraphics[scale=.5]{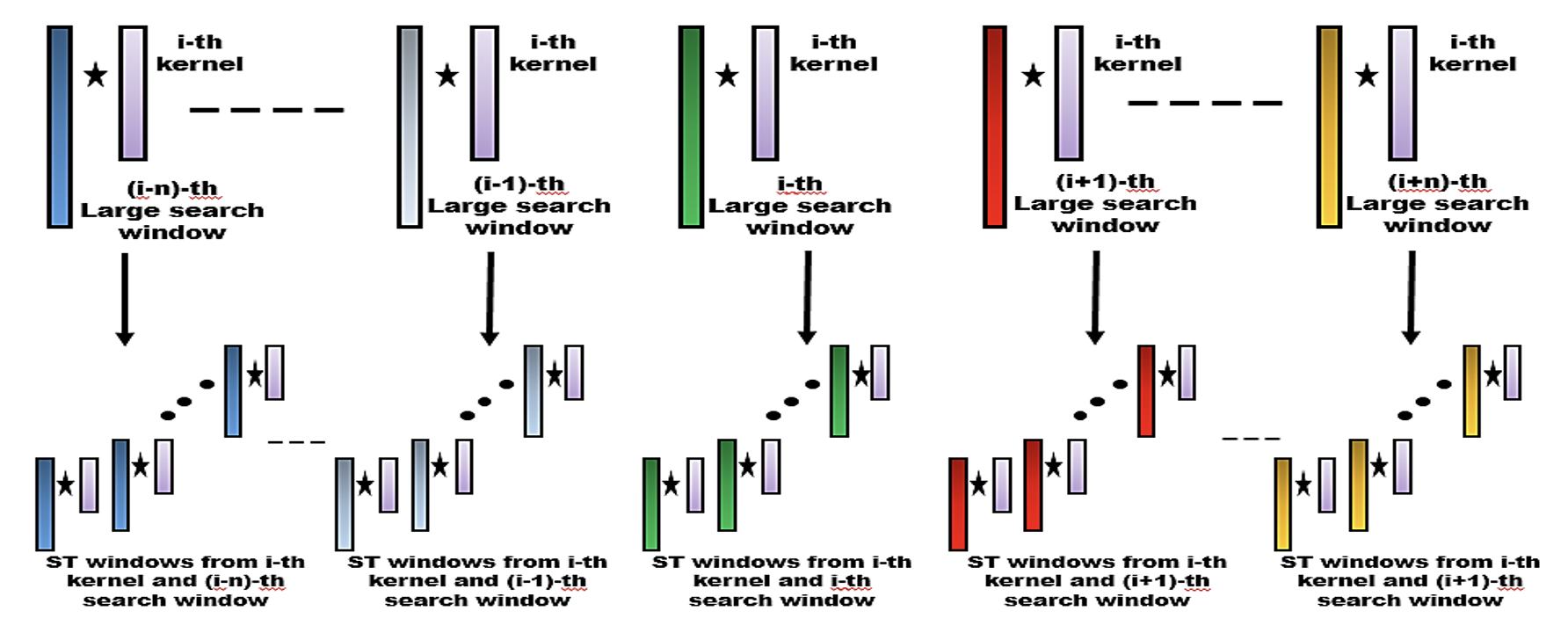}
    \caption{1.5D ST cross correlation between an array of ST kernels generated from i-th pre compression A-line with different arrays of ST search windows generated from neighboring post compression A-line in contradistinction to 1.5D cross correlation using conventional large windows. Rectangular shapes are used to denote 1D windows. Different colors denote search windows obtained from neighboring A-line RF data. }
    \label{fig:fig4}
\end{figure}
For the conventional 1.5D algorithm using large windows, a number of overlapping 1D kernels and search windows are generated from the Pre and Post compression echo signals respectively to cover each A-lines. Then, for each data point, a segment from pre-compression signal is correlated with several corresponding neighboring post-compression data segments. For example, a kernel from pre-compression i-th data stream is correlated with the corresponding search windows of (i-n)-th to (i+n)-th post-compression columns or data streams (Value of n can be varied. We have chosen n=4 for our computation based on empirical analysis of image quality and computation time). Hence a 1D kernel is correlated in a 2D search field. A total of (2n+1) cross correlation functions are generated from this process and the maximum correlation values or cross correlation co-efficient peaks are calculated from each of the correlation functions. The highest cross correlation peak among the (2n+1) values is then determined and used to locate the lateral shift. If the i-th pre-compression kernel has the highest correlation with the (i+j)-th post-compression data segment, it means that the tissue deformation took place in such a way that the corresponding post-compression segment has shifted +j columns. This lateral movement found for one segment is also used for the following segments to predict lateral movement adaptively which makes the algorithm computation efficient.
\begin{figure}[htp]
    \centering
    \includegraphics[scale=.5]{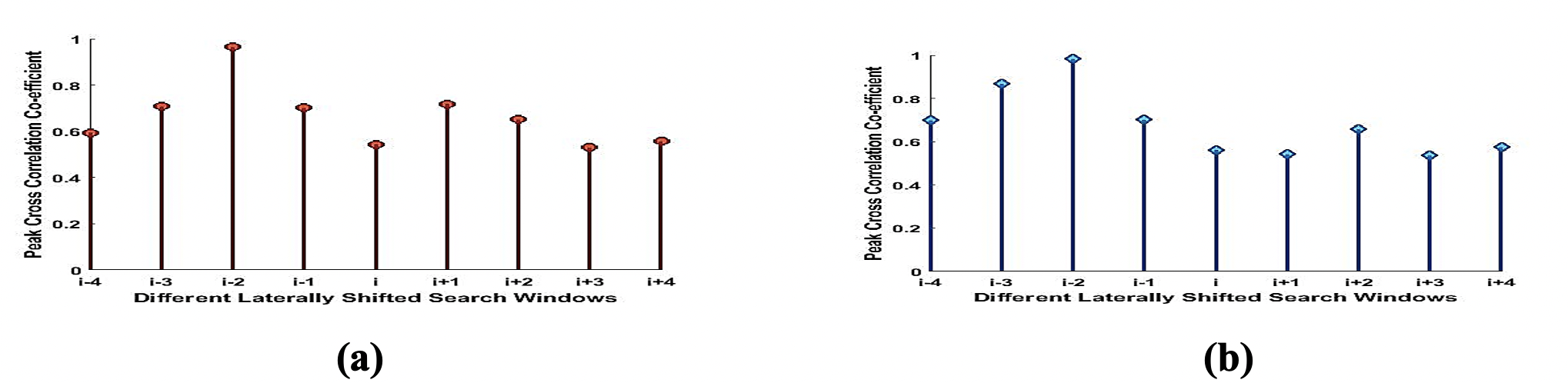}
    \caption{Highest correlation peak obtained for different lateral search window segment for    (a) Classical 1.5D algorithm and (b) 1.5D algorithm using Short Term (ST) windows    }
    \label{fig:fig5}
\end{figure}
In order to implement the 1.5D algorithm using short term windows, for each data point, instead of using a large kernel, a single array of axially overlapping short term kernels are used to cover a pre compression echo segment. Subsequently, for the post compression segment, instead of using a number of large search windows, a number of different arrays of axially overlapping short term search windows are computed from corresponding segments of laterally adjacent post compression A-lines. An array of mean cross correlation functions are generated by correlating the array of short term kernels with the different arrays of adjacent short term search windows. Additionally, each of the post compression echo frames are axially stretched adaptively using a local stretch factor that varies iteratively, to maximize the correlation between the pre and post compression windows. From the arrays of mean cross correlation functions, one with the highest cross correlation peak is elected to locate the lateral shift. The process is illustrated graphically in Figure 4 in contradistinction to conventional 1.5D algorithm with large kernels.  As the figure depicts, the same i-th kernel is correlated with (i-n)-th to (i+n)-th search windows (a total of 2n+1 search windows) to compute a lateral search. For the short term approach, instead of using conventional large windows, ST (Short Term) windows are used. An array of axially overlapping ST kernels are correlated with 2n+1 arrays of overlapping ST search windows to generate 2n+1 mean ST cross correlation functions. To validate the fruitfulness of using ST windows, we computed an experimental simulation that compares classical 1.5D algorithm using large windows and 1.5D algorithm using short term windows. The results are presented in Figure 5 and 6. First, we demonstrated the advantages of using 1.5D algorithm for 6\% applied strain. A 2.96 mm 1-D segment of correlation kernel was generated whose position is fixed to be at i-th A-line of the pre compression echo frame. Then a 2-D search window was generated by computing 1D search windows from the corresponding (i-4)-th to (i+4)-th post compression data segment. Cross correlation between the 1-D kernel and each of the search windows were computed. Then the resultant correlation peaks were computed from the respective cross correlation functions and stem-plotted in figure 5(a) from where we can derive that the correlation search window at (i-2)-th location gives the highest peak and thus demonstrates the maximum correlation with the 1D kernel. After that we generated two cross correlation functions by correlating i-th kernel with i-th search window (as we would have in classical 1-D algorithms) and i-th kernel with (i-2)-th search window and plotted them together in figure 6. We can see that, comparatively better correlation values can be obtained if we use the (i-2)-th search window instead of i-th search window. Thus it can be decoded that the post compression segment has undergone a lateral shift of -2 due to the mechanical compression and thus implementation of 1.5D algorithm allows us to locate the correct lateral shift and obtain better elastograms. To explore the effect of using short term windows instead of conventional large ones in 1.5D algorithm, we computed an array of short term kernels from the pre-compression echo frame with each having a length of 0.74 mm and 75\% axial overlapping along with 9 arrays (n=4) of axially overlapping short term search windows from laterally neighboring A-lines of the post compression echo frame. The array of short term kernels were correlated with each of the 9 arrays of short term search windows and the 9 resultant mean cross correlation functions were computed. The highest correlation peaks were calculated from the mean cross correlation functions and stem-plotted in figure 5(b). From figure 5(b) we can observe that the highest correlation is obtained for (i-2)-th array of short term search windows, which is similar to the conventional 1.5D algorithm using large windows. However, the value of correlation peaks have increased slightly after implementation of short term windows. The highest correlation value obtained for conventional 1.5 D algorithm is 0.9670 while the highest obtained correlation peak for the 1.5D algorithm using short term windows is 0.9848. In order to evaluate the significance of this increment, we can use the following equation where the Signal to noise ratio (SNR) is expressed in terms of cross correlation co-efficient ~\cite{585223}:

\begin{equation}
S N R=\frac{\rho_{x y(t o)}}{1-\rho_{x y(t o)}}
\end{equation}

Where $\rho_{xy(to)}$ is the cross correlation co-efficient between the pre and post compression signals at the time delay of to. Using equation 9 if we compute the SNR value based on the cross correlation peak obtained for the conventional 1.5D algorithm and 1.5D algorithm using short term windows, then the SNR value for the 1.5D method using large windows will be 29.30 and the SNR for the 1.5D approach using ST windows will be 64.79. Hence, even though the cross correlation peak increased by 0.0178 after implementing the ST windows, it shows the potential to increase the SNR by 54.78\%. 
\begin{figure}[htp]
    \centering
    \includegraphics[scale=.5]{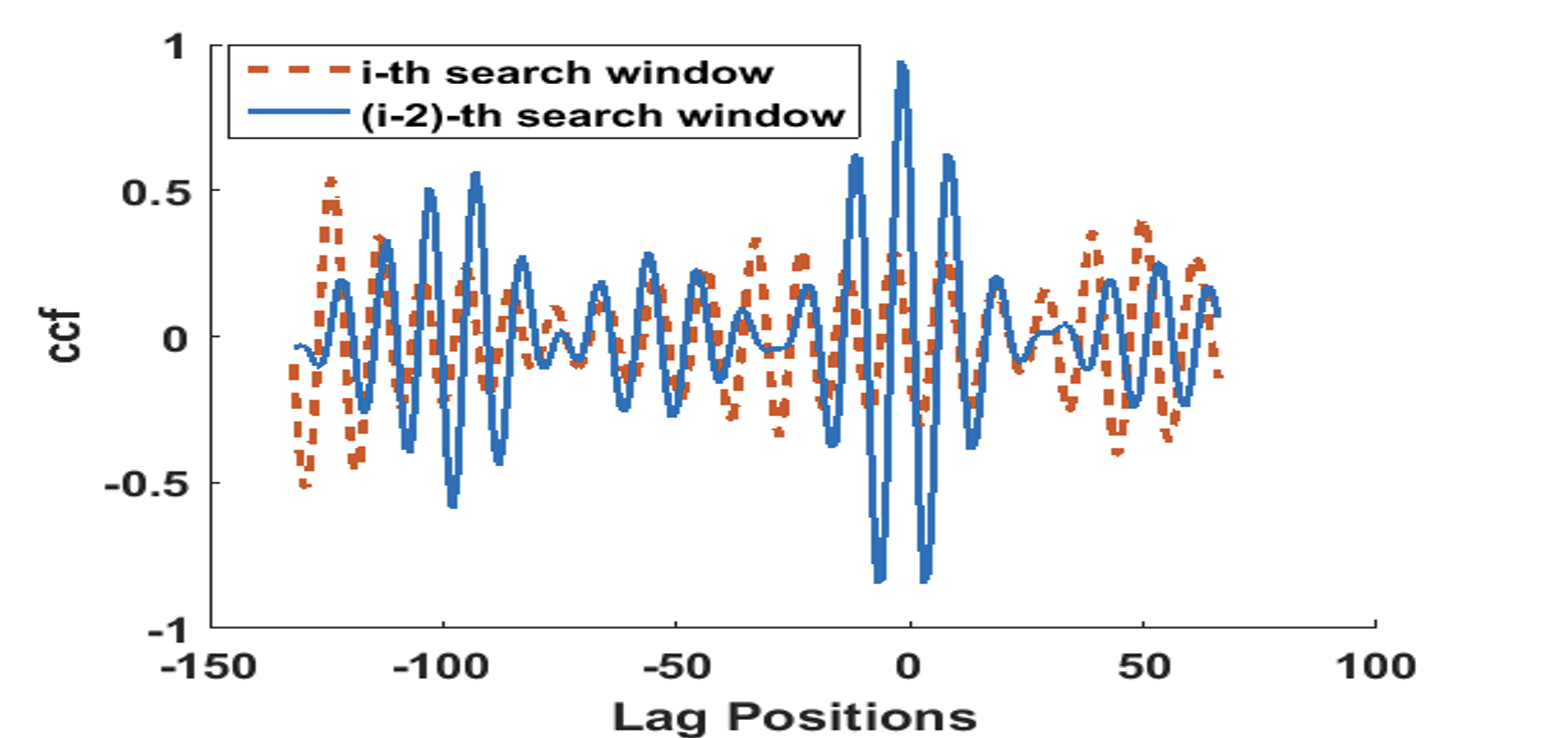}
    \caption{Cross Correlation Co-efficient (ccf) vs Lag Positions. Correlation using correct vs incorrect search window  }
    \label{fig:fig6}
\end{figure}

\subsection{ Quantitative Parameters for performance evaluation:}

In order to evaluate the performance of our proposed algorithm, we performed a rigorous assessment using both simulation and in-vivo patient data and presented an extensive comparative validation. We analyzed the strain maps generated by different algorithms using visual inspection and also performed a quantitative inspection using two numerical parameters, namely, Strain Signal to Noise ratio (SNRe) and Strain Contrast to Noise ratio (CNRe). The two quantitative metrics are defined as follows:

(1) Strain Signal to Noise Ratio (SNRe) ~\cite{Cespedes1993-gl}: 

\begin{equation}
S N R e=\frac{\mu_S}{\sigma_S}
\end{equation}

In Equation 10, $\mu_s$ and $\sigma_s$ respectively implies the statistical mean and standard deviation computed from a homogeneous location in the strain map.

(2) Strain Contrast to Noise Ratio (CNRe)~\cite{Varghese1998-fj}: 

\begin{equation}
C N R e=\frac{2\left(\mu_l-\mu_b\right)^2}{\sigma_l^2+\sigma_b^2}
\end{equation}

In Equation 11, $\mu_l$ is the mean strain computed from an approximately homogeneous strain region inside the lesion and $\mu_b$  represents the same computed from an approximately homogeneous background tissue region. Whereas, the symbols $\sigma_l$ and $\sigma_b$ represent strain standard deviation computed from approximate homogeneous regions inside the lesion and tissue background respectively.

The two terms are expressed in decibel scale using the following logarithmic conversion:

\begin{equation}
X_{d B}=20 \log _{10}(X)
\end{equation}

Where X denotes SNRe or CNRe.

\subsection{Search Algorithm:}

The search algorithm devised to compute strain maps based on our proposed method is illustrated in Figure 7(a). The colored box represents the method selected to elect the next value of stretch factor $\alpha$, which is the Binary search method shown in figure 7(b). The stretch factor must be iteratively varied until the value that can maximize the correlation between pre and post compression signal is reached. For our our work, we have implemented the Binary search method which is computationally intensive. The binary search method is shown in figure 7(b) where an initially chosen search interval is halved after each iteration while retaining the stretch factor that gives higher correlation.   
To generate the strain maps, the entire processing software was coded in MATLABTM (The Mathworks, Inc., Natick, MA). For gamma correction and gray scale mapping of in-vivo elastograms, Registax 6 (Developed by Cor Berrevoets, The Netherlands) was utilized.  

\subsection{Experimental Metods and Materials:}
\subsubsection{2D Finite Element Model Phantom Simulation Data: }

The simulation phantom used for this analysis is a two dimensional Finite Element model (FEM) which was generated by using the analysis software ALGOR (ALGOR is a registered trademark of Algor, Inc). It had a 40 mm X 40 mm rectangular proportion with a homogeneous background and contained four circular inclusions with each having a diameter of 7.5 mm. The background stiffness was configured to be 60 kPa which is akin to the mean stiffness of typical breast glandular tissue. Each of the circular inclusions had different stiffness as illustrated in Figure 8(a). The top, middle, bottom right and bottom left lesions were 20 times (20 DB), 100 times (40 DB), 31.62 (30 DB) and 3.12 times (10 DB) stiffer than the homogeneous background respectively. A total of 30372 scatterers were simulated in the phantom along with sonographic SNR of 40 dB by integrating random white noise. While constraining the bottom part vertically, the phantom was compressed at the top using a compressor that had width larger than that of the phantom. And stiffness was configured to be 60 kPa which is akin to the mean stiffness of typical breast glandular tissue. Each of the circular inclusions had different stiffness as illustrated in Figure 8(a). The top, middle, bottom right and bottom left lesions were 20 times (20 DB), 100 times (40 DB), 31.62 (30 DB) and 3.12 times (10 DB) stiffer than the homogeneous background respectively. A total of 30372 scatterers were simulated in the phantom along with sonographic SNR of 40 dB by integrating random white noise. While constraining the bottom part vertically, the phantom was compressed at the top using a compressor that had width larger than that of the phantom.

\begin{figure}[htp]
    \centering
    \includegraphics[scale=.6]{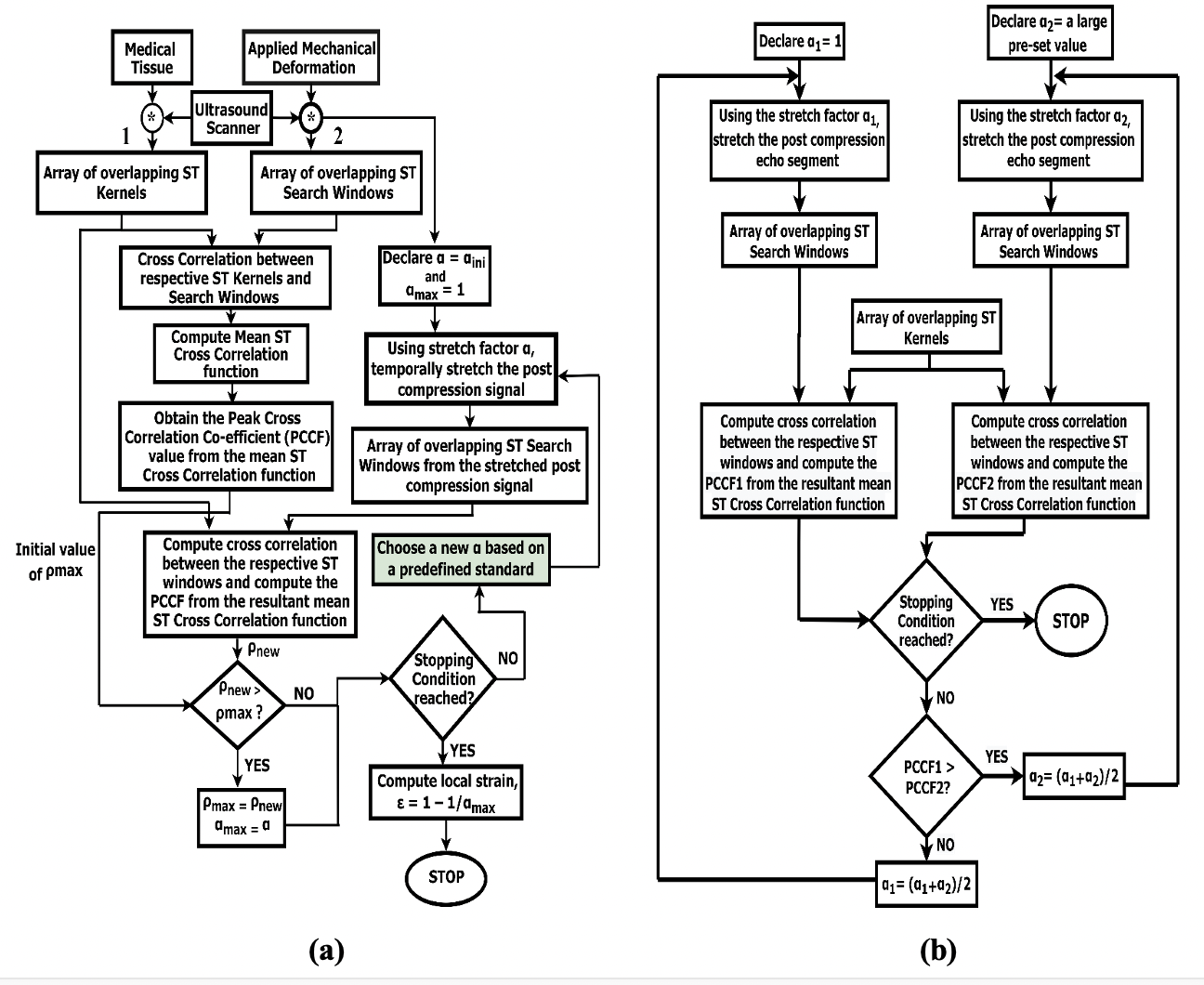}
    \caption{(a) Algorithm Flowchart. Here "*" represents convultion. The numeric notation ‘1’ denotes pre-compression signal and ‘2’ represents post-compression signal (b) Binary Search method to find the next value of $\alpha$}
    \label{fig:fig7}
\end{figure}

An ultrasonic transducer having specification of 5 Mhz center frequency and fractional bandwidth of 60\% was used to scan the phantom from the top by employing a non-diffracting Gaussian beam with 1.5 mm diameter. The phantom was allowed to expand freely on both the axial direction (top and bottom) and a total of 128 scan lines were obtained. The ideal strain map of FEM simulated data is illustrated in figure 8(b). Strain Elastograms are generated at applied strain of 2\%, 4\%, 6\% and 8\% by using different algorithms and are shown in figure 9 for a qualitative assessment. For data processing using classical adaptive stretching and adaptive stretching with lateral search, we maintained the length of conventional large correlation kernel to be 2.96 mm. whereas, for our proposed short term algorithms, 0.74 mm short term kernels with 75\% axial overlapping was maintained. Median filtering was used only on the computed elastograms to remove shot noise and must not be confused with any form of correlation filter.

\subsubsection{In-vivo Data of breast tumor:}
We performed another comparative analysis using in-vivo (processes performed or taking place in living organisms) patient data of breast tumors elected from a database consisting 47 cases in total. These patients were within the age limit of 20 to 75 years and gave their consent for being a part of this study. The database comprises of both Benign and Malignant tumors. The study was approved by the Institutional Review Board (IRB). An expert clinician conducted the medical examination and for data acquisition a SONIX-500 RP (Ultrasonix Medical Corporation, Richmond, BC, Canada) commercial ultrasound scanner combined with an L14-5/38 probe (which was operating at a nominal value of 10 Mhz) was used. This study was conducted at the University of Vermont Medical Center, Burlington, VT, USA (UVM). While acquiring the data, the clinician performed free hand compression while the patients were positioned in a supine of oblique supine position so that the breast was evenly flattened against the chest wall. Resultant strain maps and B-mode images were observed and analyzed from a cine clip of three cycles of compression-relaxation containing b-mode images, strain images and corresponding RF data.  The rf data that we have used to generate our strain images came from the frames corresponding to the highest quality strain images recorded in the cine clip.  Other useful information such as size and position of the lesion were also recorded. The histopathology reports were carefully cached from where the ground truth was obtained. For this study, one malignant tumor (Adenocarcinoma) and one benign tumor (Fibroadenoma) were used to analyze the performance of our proposed method.
%%%%%%%%%%%%%%%%%%%%%%%%%%%%%%%%%%%%%%%%%%%%%%%%%%%%%%%%%%%%%%%%%%%%%%%%%%%%%%%%%%%%%%%%%%%%%%%%%%%%%%%%%%%%%%%%%%%%%%%%%%%%%%%%%%%%%%%%%%%%%%%%%%%%%%
\section{Results}
\subsection{Validation Using Simulation Data:}

\begin{figure}[htp]
    \centering
    \includegraphics[scale=.3]{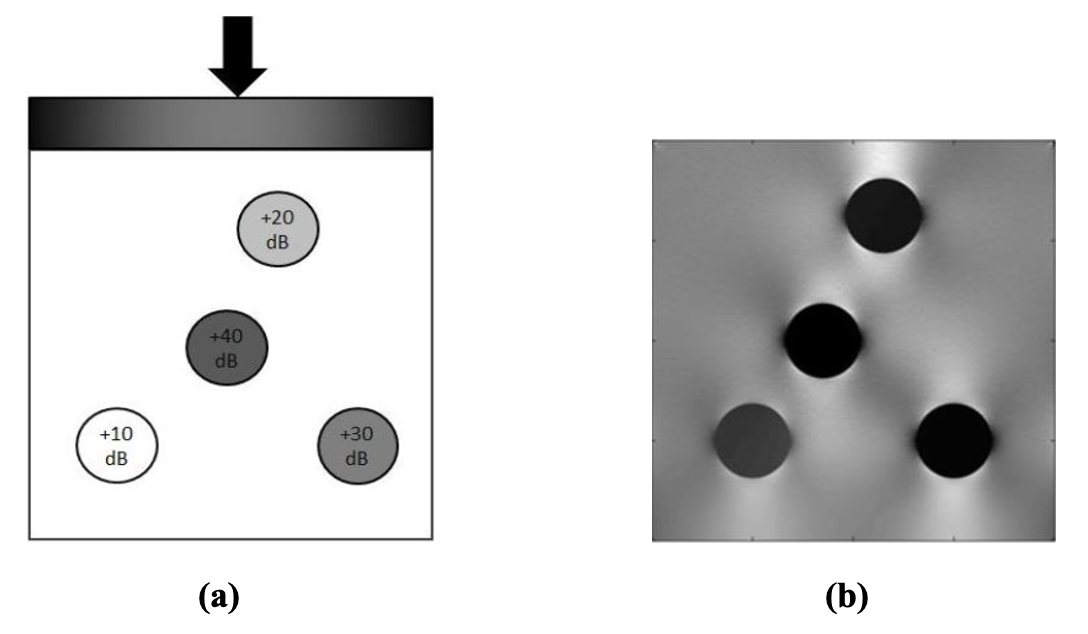}
    \caption{Finite Element Model simulation phantom. (a) Inclusions with different levels of stiffness compared to a homogeneous background of 60 kPa, (b) Ideal Strain Elastogram of the corresponding simulation Phantom }
    \label{fig:fig8}
\end{figure}
Strain images generated by using the classical adaptive stretching algorithm are shown in figure 9(a). At lower levels of applied strain, the strain images are free from any artifacts having lesser background noise and the lesions are clearly visible. The bottom two lesions (10dB and 30 dB) show marginal irregularities in their shape and boundary definition. At high compression levels (6\% and 8\%) however, background noise can be observed, mostly along the lateral direction from the image center and at 8\% strain, the strain map suffers from significant distortion as the bottom two lesion appears to be blurry and surrounded by considerable amount of background noise. This can be explained from the fact that, at higher strain levels, more lateral shift is induced which compounds the decorrelation noise and lead to poor strain maps. Figure 9(b) shows the strain elastograms generated by using adaptive stretching with short term windows. These strain maps are evidently less noisy and much closer to the ideal elastogram compared to the images in Figure 9(a). Although they are not completely free from background noise at higher strain levels, the lesions are better depicted compared to the classical approach. The effect of laterally induced decorrelation noise is significantly lower in the strain elastograms shown in Figure 9(c). These images are generated by using adaptive stretching with the 1.5D lateral search. At higher strain levels, the background appears to be significantly smooth free from distortion noise. However, the lesions show irregular and distorted shape for the 6\% and 8\% applied strain. The elastograms in Figure 9(d) clearly appears to be better than any other set of strain maps. These images are generated by implementing Adaptive Stretching algorithm with short term kernels along with the lateral search. All the images are free from any artifacts or background noise and the lesion boundaries are clearly depicted even at higher levels of applied strain. Even though marginal irregularities can be noted in the 40 dB lesion’s configuration at 8\% strain, the strain maps in figure 9(d) are comparatively much better than the other images and closely resembles the ideal elastogram.

\begin{figure}[htp]
    \centering
    \includegraphics[scale=.6]{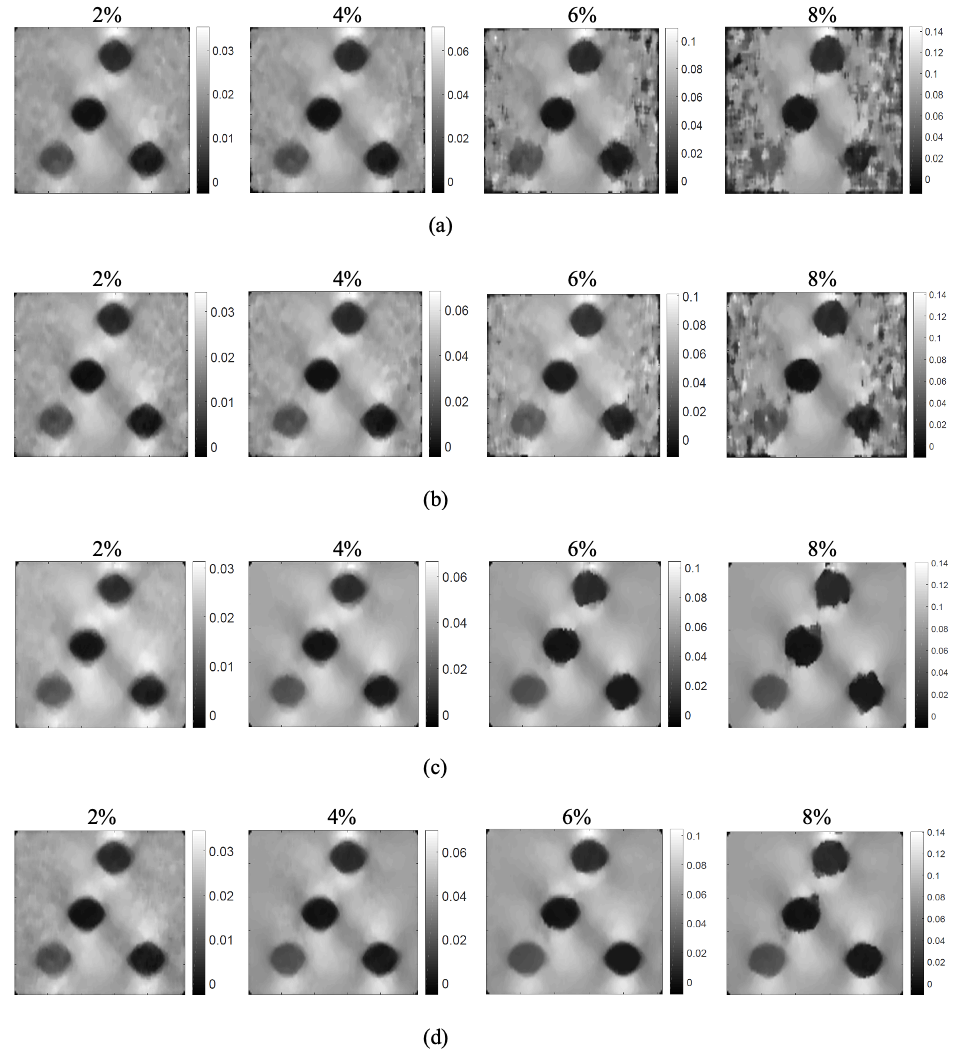}
    \caption{Strain Elastograms of the simulated phantom computed by using different algorithms at applied strain of 2\%, 4\%, 6\% and 8\% (a) Adaptive Stretching (AS) (b) Adaptive Stretching using Short Term Windows (AST) (c) Adaptive Stretching with Lateral Search (ASL) (d) Adaptive Stretching with Lateral Search using Short Term Windows (ASLST)  }
    \label{fig:fig9}
\end{figure}

To evaluate the performance of our proposed method from a quantitative point of view, the two numerical parameters in equations 10 and 11 are calculated from strain elastograms generated by different algorithms at varying levels of applied strains (1-8\%). The results are presented graphically in Figure 10.

Figure 10 (a)-(d) exhibits the CNRe plots for each of the four inclusions with varying stiffness. The CNRe plots reveals that for classical adaptive stretching (AS) the performance degrades as the level of applied strain increases. Adaptive stretching with short term windows (AST) outperforms the classical method considerably at higher strain values while the difference is marginal for lower strain levels. However, for 10 dB and 30 dB lesion, the CNRe values decreases sharply at 8\% strain. Adaptive stretching with lateral search (ASL) performs better at higher strain values than the 1-D methods albeit a sharp decline is noted for the 30 dB lesion. Highest values of strain CNRe can be observed for Adaptive Stretching with lateral search using short term windows (ASLST) as it eclipses all three algorithms for all the different strain values. CNRe values for ASL algorithm are in close proximity with ASLST in case of 20 dB and 10 dB lesions but the ASLST method surmounts the ASL method by significant margin for the 40 dB and 30 dB lesions.  Similar trend can also be noticed from Figure 10(e) which illustrates the SNRe plot. The SNRe values reveals that ASLST triumphs the ASL method while the AST algorithm outperforms the classical AS method by a significant margin.

\begin{figure}[htp]
    \centering
    \includegraphics[scale=.5]{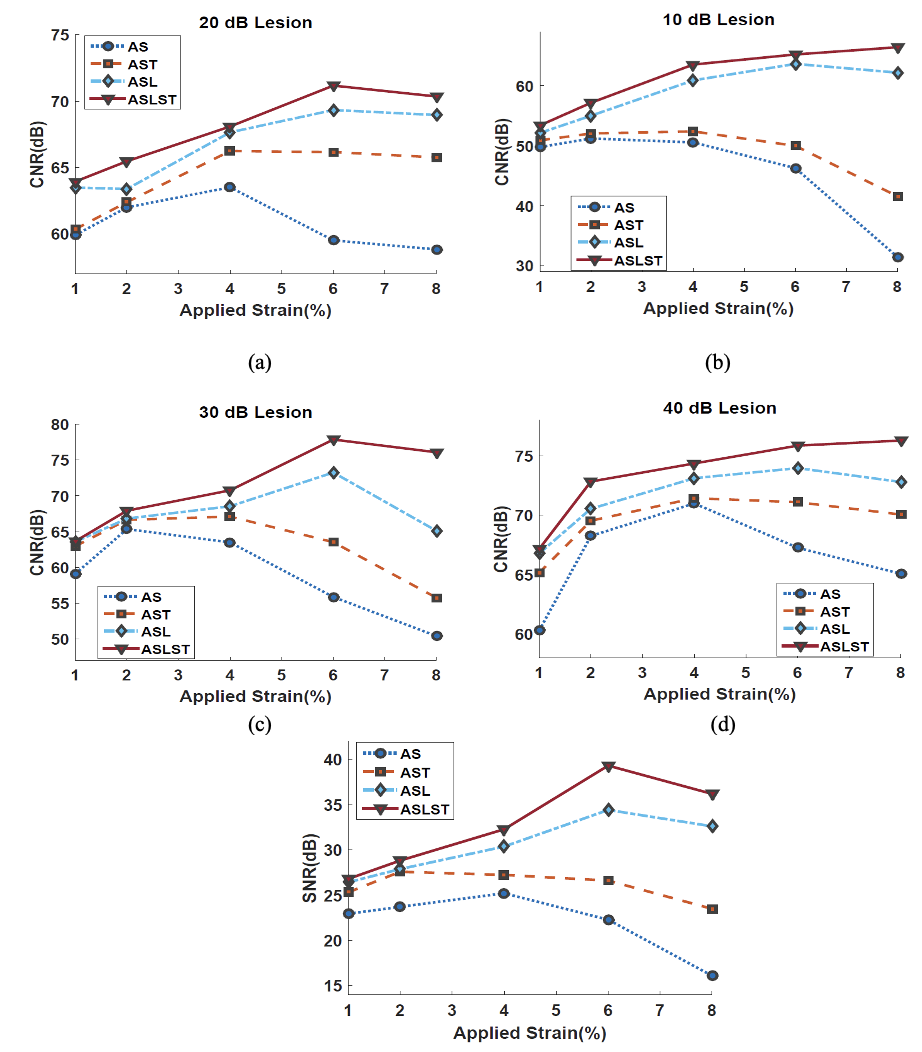}
    \caption{Strain CNR for different lesions (a)20 Db, (b)10 dB, (c)30 dB, (d)40 dB along with         (e) Strain SNR for different algorithms.  }
    \label{fig:fig10}
\end{figure}

\subsection{Validation Using In-vivo data:}
The strain maps generated for the malignant and benign tumors by using different algorithms are shown in Figure 11 and 12 respectively. In order to have a higher contrast view of the interior of the lesion, the gamma functions of the strain images were adjusted by a gray scale mapping and presented along with the original elastograms. The B-mode images are also provided along with the strain maps. For visual inspection, one key feature that must be noted is that, in strain elastograms, most malignant breast tumors are expected to be larger in size compared to the hypoechoic portion of the lesion on corresponding ultrasonic B-mode images while the benign tumors appear to be smaller or equal in size [33],[34]. From Figure 11 we can see that for all four methods the malignant tumor appears to be bigger in size compared to the lesion in B-mode image. However, the strain maps generated by Adaptive Stretching (AS) in Figure 11(a) bears considerable background noise and the lesion boundary is somewhat irregular. The strain map in Figure 11(b) is generated by implementing Adaptive Stretching with short term windows (AST) and it carries better result compared to Figure 11(a) as the elastogram appears to be smoother and the amount of surrounding noise around the lesion is quite less. The lesion boundary is also better depicted. Figure 11(d), which shows the strain image generated by Adaptive Stretching with 1.5D lateral search using ST windows (ASLST), holds comparatively the best strain elastogram which shows the least amount of background noise and the lesion is better depicted with higher contrast and sharper boundary. Figure 12 shows the strain maps generated for the benign tumor. In each case, the lesion size in strain elastogram is somewhat similar or marginally smaller than the lesion in B-mode image. Figure 12(b) and 12(d) shows the strain images generated by using short term windows and they appear to be better than the strain images in Figure 12(a) which is generated by the classical AS method. Figure 12(c) illustrates the strain elastogram computed by using the ASL method. Although it appears to have lesser noise compared to figure 12(b), the boundary edges appear to be sharper for the other 3 methods. 

\begin{figure}[htp]
    \centering
    \includegraphics[scale=.5]{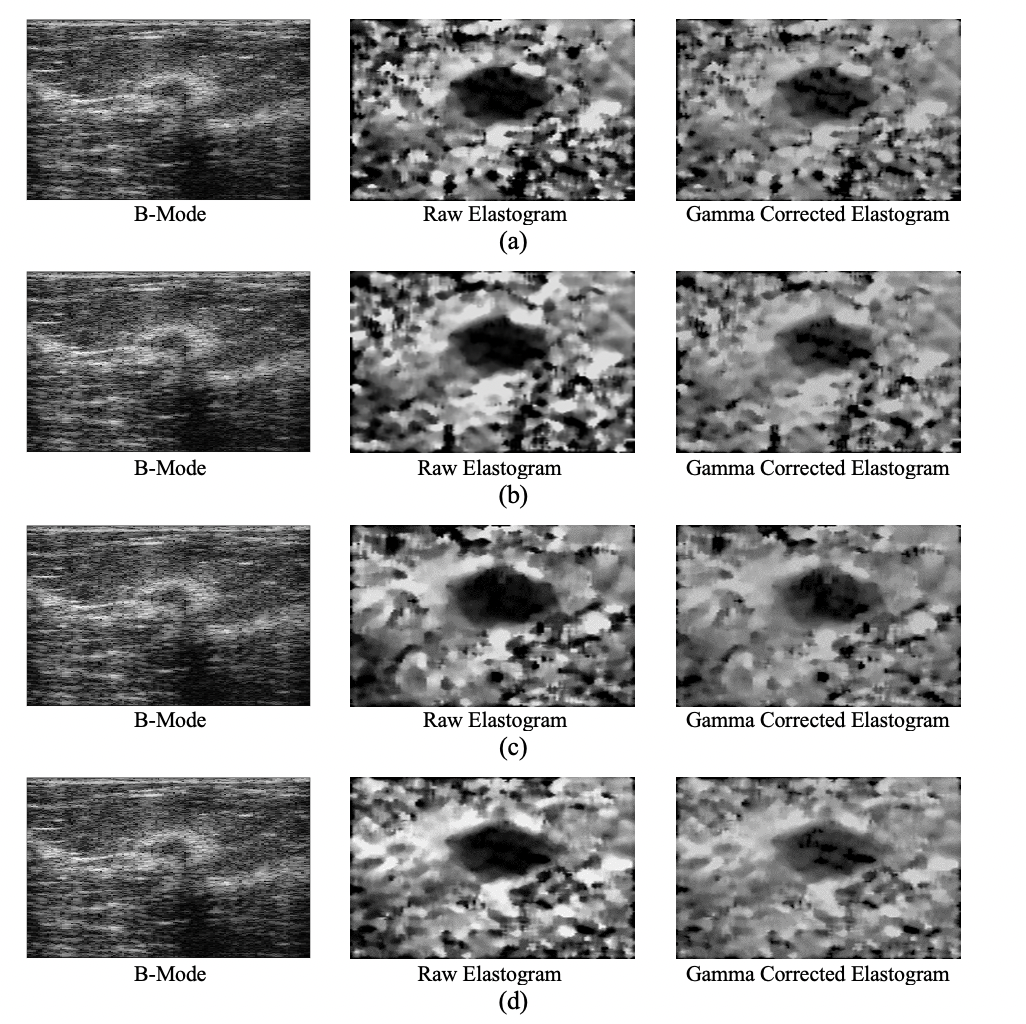}
    \caption{Strain Elastograms of an in-vivo Malognant tumor computed by employing different algorithms (a) Adaptive Stretching (AS) (b) Adaptive Stretching using Short Term windows (AST) (c) Adaptive Stretching with Lateral Search (ASL) (d) Adaptive Stretching with Lateral Search using Short Term windows (ASLST)  }
    \label{fig:fig11}
\end{figure}

For a quantitative assessment, the SNR and CNR values are computed from the strain images in figure 11 and 12 using equation 10 and 11 respectively and presented in table 1 and 2. The highest SNR and CNR values are obtained by using the ASLST method which are much higher than the conventional 1D AS method. Using ST windows as a substitute to large windows lead to higher SNR and CNR values. At the same time, 1.5D Algorithms performed better compared to the 1D algorithms in terms of these quantitative features. The improvement in performance is more significant in malignant tumor compared to the benign one. 

\begin{figure}[htp]
    \centering
    \includegraphics[scale=.5]{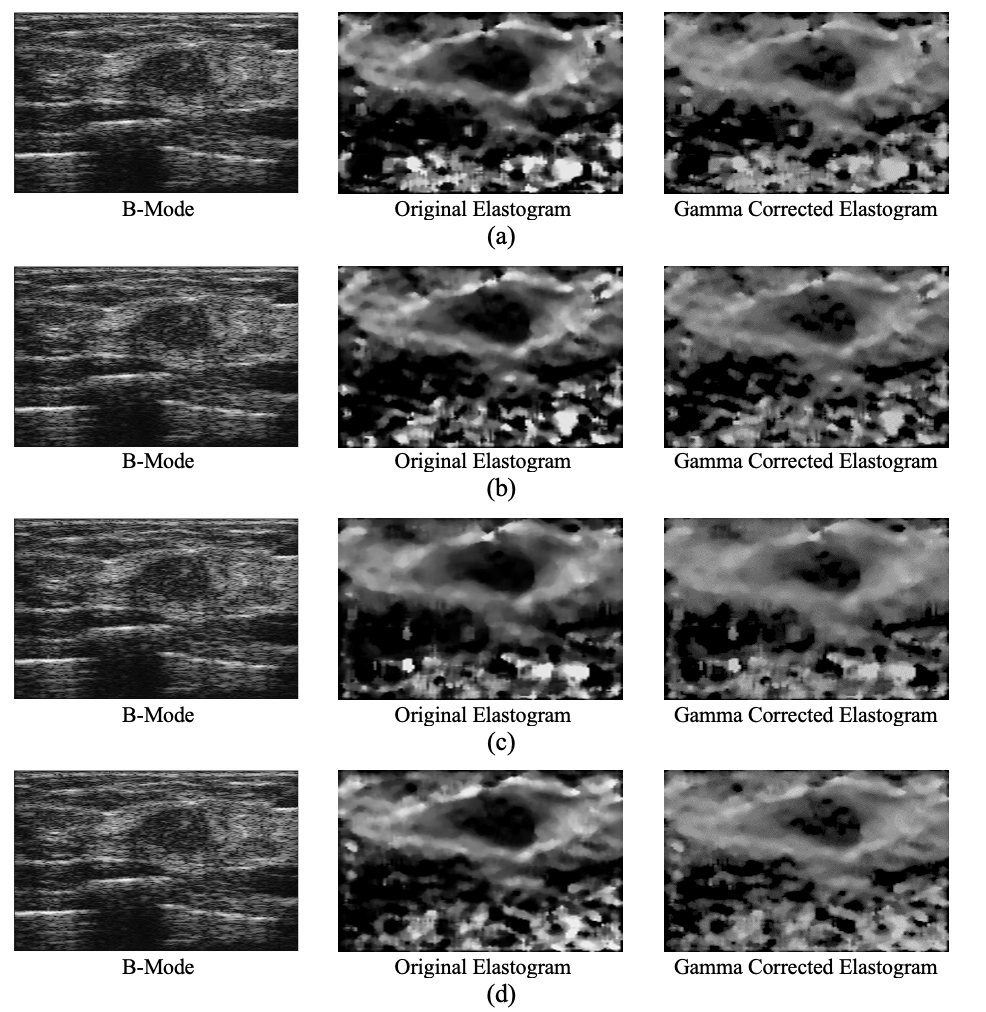}
    \caption{Strain Elastograms of an in-vivo Beningn tumor computed by employing different algorithms (a) Adaptive Stretching (AS) (b) Adaptive Stretching using Short Term windows (AST) (c) Adaptive Stretching with Lateral Search (ASL) (d) Adaptive Stretching with Lateral Search using Short Term windows (ASLST) }
    \label{fig:fig12}
\end{figure}

In order to quantitatively examine the sharpness of the strain images along the tumor boundary, magnitude of the image intensity gradient function was estimated from multiple 2.1 mm x 1.5 mm regions along the tumor boundary. The mean gradient value was then obtained and utilized for comparative assessment of the strain images generated by different algorithms as depicted in table 3. The Adaptive Stretching algorithm using ST windows (AST) generated strain images with sharped tumor boundary compared to the conventional Adaptive Stretching (AS) method using large windows. The tumor boundary in the strain maps computed by the conventional 1.5D algorithm (ASL) were less sharp compared to the 1D algorithms. However, when the 1.5D method was implemented using ST windows (ASLST) the sharpness of the tumor boundary increased significantly for both the benign and malignant tumors and this algorithm performed much better than the other three methods in terms of boundary sharpness.

\begin{table}[]
\centering
\caption{SNR of the benign and malignant tumors computed by different algorithms}
\label{tab:comparison}
\begin{tabular}{|l|l|l|}
\hline
Method & SNR(dB) for Benign & SNR(dB) for Malignant \\ \hline
Adaptive Stretching (AS)    & 14.41	&16.25 \\ \hline
Adaptive Stretching using ST windows (AST)    & 16.53 &	17.33 \\ \hline
Adaptive Stretching with 1.5D lateral search (ASL)    & 17.81 &	19.44 \\ \hline
Adaptive Stretching with 1.5D lateral search using ST windows (ASLST)    & 18.44 & 21.27 \\ \hline

\end{tabular}
\end{table}

\begin{table}[]
\centering
\caption{CNR of the benign and malignant tumors computed by different algorithms}
\label{tab:comparison}
\begin{tabular}{|l|l|l|}
\hline
Method & SNR(dB) for Benign & SNR(dB) for Malignant \\ \hline
Adaptive Stretching (AS)    & 36.97	& 40.95 \\ \hline
Adaptive Stretching using ST windows (AST)    & 41.90 &	43.21 \\ \hline
Adaptive Stretching with 1.5D lateral search (ASL)    & 44.01 &	47.37 \\ \hline
Adaptive Stretching with 1.5D lateral search using ST windows (ASLST)    & 46.77 &	52.89 \\ \hline

\end{tabular}
\end{table}

\begin{table}[]
\centering
\caption{Mean Gradient }
\label{tab:comparison}
\begin{tabular}{|l|l|l|}
\hline
Method & SNR(dB) for Benign & SNR(dB) for Malignant \\ \hline
Adaptive Stretching (AS)    & 0.8373 &	1.2688 \\ \hline
Adaptive Stretching using ST windows (AST)    & 0.9174 &	1.4063 \\ \hline
Adaptive Stretching with 1.5D lateral search (ASL)    & 0.8053 &	1.0366 \\ \hline
Adaptive Stretching with 1.5D lateral search using ST windows (ASLST)    & 1.0346 &	1.5231 \\ \hline

\end{tabular}
\end{table}

\section{conclusion}
We presented a novel strain estimator that implements adaptive strain estimation algorithm using short term correlation kernels. The effects of using short term windows instead of conventional large ones have been thoroughly investigated using simulation phantom and in-vivo data of breast lesions which corroborated the effectiveness of our proposed method. To account for the lateral shift, a lateral search has been integrated that employs a 1.5D algorithm. Short term correlation kernels are effective in reducing decorrelation noise and generating strain elastograms with apparent higher resolution. Peak hopping or False Peaks are major sources of error while using short term correlation as they can lead to erroneous displacement estimates. To eliminate false peaks, we have simply computed a mean short term correlation function from which the location of the true peak could be accurately tracked. This eliminates the implementation of additional correlation filter. Designing a correlation filter can be computationally intensive as its length must be configured accurately. To estimate strain we have implemented a more direct adaptive approach which computes strain from a local stretch factor of the post compression signal. The stretch factor is iteratively varied to maximize correlation between the pre and stretched post deformation echo frames. This method is suitable for the heterogeneous nature of biological tissues as it can track the non-uniform displacement with greater precision. But non-axial motion still occurs and to account for this, a lateral search must be incorporated. In this work, we have implemented a lateral search by correlating an array of 1-D short term correlation kernels and different arrays of short term search windows computed from laterally neighboring post compression A-lines. This method utilizes 1-D correlation in a 2-D search location. 
One disadvantage of our proposed method is its higher computation time compared to classical adaptive stretching. However, with modern technologies computation time has become less of an issue and hence the tradeoff between computation time and higher computational efficiency and high SNRe strain elastograms can be accepted. 
The effect of using spectral parameters as strain estimators using short term windows can be investigated in further study of our proposed method. Adaptive spectral estimators however have previously appeared to be more robust but less accurate than temporal strain estimators [18]. The compound effect of adaptive strain estimators and early spectral estimation in the form of adaptive frequency scaling can be implemented and analyzed to enhance the performance of adaptive strain estimation using short term windows.   
To conclude, our proposed method can generate strain images with higher SNRe compared to conventional adaptive stretching and the elastograms demonstrate better image contrast and resolution. Also, lesions can be depicted accurately at high applied strain with sharper boundary definition which would facilitate the clinicians to detect and demarcate the tumors more precisely

\bibliographystyle{IEEEtran}
\bibliography{main}

\end{document}